\pdfoutput=1
\documentclass[11pt]{iopart}
\usepackage{iopams}  
\usepackage[utf8]{inputenc}

\usepackage[square,numbers,sort&compress]{natbib}
\expandafter\let\csname equation*\endcsname\relax
\expandafter\let\csname endequation*\endcsname\relax

\usepackage[usenames]{color}
\usepackage{commath}
\usepackage[varg]{txfonts}
\usepackage{graphicx}
\usepackage{longtable}
\usepackage{hyperref}
\usepackage{tcolorbox}
\usepackage{siunitx}
\usepackage[version=3]{mhchem}
\usepackage[displaymath]{lineno}

\definecolor{AL}{rgb}{1.0,0,0}

\begin{document}

\title{Modeling streamer discharges as advancing imperfect conductors}
\date{\today}
\submitto{Plasma Sour. Sci. Tech.}
\author{A. Luque, M. González, F.J. Gordillo-Vázquez}
\address{Instituto de Astrof\'isica de Andaluc\'ia (IAA), CSIC, Granada, Spain}
\ead{aluque@iaa.es}

\vspace{10pt}

\vspace{10pt}
\begin{indented}
\item[]\today
\end{indented}

\begin{abstract}
  A major obstacle for the understanding of long electrical discharges is 
the complex dynamics of streamer coronas, formed by many thin conducting 
filaments.  Building macroscopic models for these filaments is one
approach to attain a deeper knowledge of the discharge corona.  
Here we present a one-dimensional, macroscopic model of 
a propagating streamer channel.  We represent the streamer
as an advancing finite-conductivity channel with a surface charge density 
at its boundary.  This charge evolves self-consistently due to the electric 
current that flows through the streamer body and within a thin layer at 
its surface. We couple this electrodynamic evolution with a field-dependent 
set of chemical reactions that determine the internal channel conductivity.  
With this one-dimensional model we investigate how key properties of 
a streamer affect the channel's evolution.  The ultimate objective of our
model is to construct realistic models of streamer coronas in order to
understand better the physics of long electrical discharges.
\end{abstract}

%

\section{Introduction}
\label{sect:introduction}
Appearing often as the initial stage of a gas discharge, 
a streamer is an ionized 
filament that advances due to electron impact ionization at its tip.
Typically tens to hundreds of streamers emerge from a pointed electrode
after the sudden application of an intense electric field.  Streamers are also
the building blocks of high-altitude discharges in our atmosphere and they
precede and drive the propagation of hot leader channels in long gaps and
in lightning.

Although the microphysics of a streamer is now relatively well understood, we 
still lack solid macroscopic models to understand the long-time properties
of a streamer channel and the interactions between all filaments within a large
streamer corona.  These two issues appear to be particulary important in 
relation to the streamer-to-leader transition, in which sections of a streamer 
corona are heated up to temperatures of a few thousand Kelvin where thermal
ionization becomes significant.

Our lack of macroscopic models is particularly aggravating since at a coarse
level streamers appear to be essentially one-dimensional objects; one expects
(or rather wishes) that they can be modelled by abstracting away microscopic 
details and considering only macroscopic quantities such as the channel width, 
the linear charge density and the tip velocity.  This was the motivation for 
the model for streamer trees presented in ref.~\citep{Luque2014/NJPh}, where
the macroscopic dynamics were justified 
in part by phenomenological considerations and in part by appealing 
to experimental data.  For example, the electrostatic interaction between
different channel segments was modelled by an ad-hoc kernel derived as 
the simplest expression that satisfies some required properties.
The electrical conductivity of the channel was also fixed and not calculated 
self-consistently.

In this article we build a more detailed one-dimensional
model where a streamer is described as an imperfect conductor that grows 
within an external field.  Our purpose here is not to derive quantitative
properties of actual streamers but rather to investigate
the relations between macroscopic quantities.  By directly controlling some
magnitudes such as streamer velocity, which in microscopic models emerge as 
derived properties, we can answer questions such as how the peak 
electric field in a streamer depends on its velocity. 

Some other approaches have been developed to simplify the problem of streamer
propagation.  Lozanskii \cite{Lozanskii1975/SvPhU} proposed to consider the
streamer interior as a perfect conductor and thus the streamer boundary as
an equi-potential surface.  Moving-boundary (also called contour-dynamics)
methods \cite{Kao2010/PhysD,Arrayas2010/PhRvE,Ebert2011/Nonli/1}
derive from this approach and have been applied to investigate Laplacian
branching of streamers 
\cite{Meulenbroek2004/PhRvE,Meulenbroek2005/PhRvL,Arrayas2012/PhRvE}
and the role of streamer curvature \cite{Brau2008/PhRvE/1,Brau2009/PhRvE}.
Recently these models have also incorporated a finite internal conductivity
\cite{Brau2008PhRvE/2,Arrayas2011/PhRvE} but they are generally limited to
short streamers and relatively simple settings such as homogeneous 
background fields.  Another family of reduced streamer models derives
from the Dielectric Breakdown Model first proposed
by Niemeyer and coauthors \cite{Niemeyer1984/PhRvL}.  In these models a streamer
corona expands stochastically by the random accretion of filaments with a
field-dependent probability.  A variation of this model was applied to 
sprite discharges in the mesosphere \cite{Pasko2000/GeoRL}.  Finally we mention
corona models such as the one developed by Akyuz \cite{Akyuz2003/JElec}, which
considered a branched tree of several perfectly-conducting channels.

\section{Model}
\label{sect:model}
\subsection{Charge transport}
\label{sect:current}
\begin{figure}
\includegraphics[width=0.9\columnwidth]{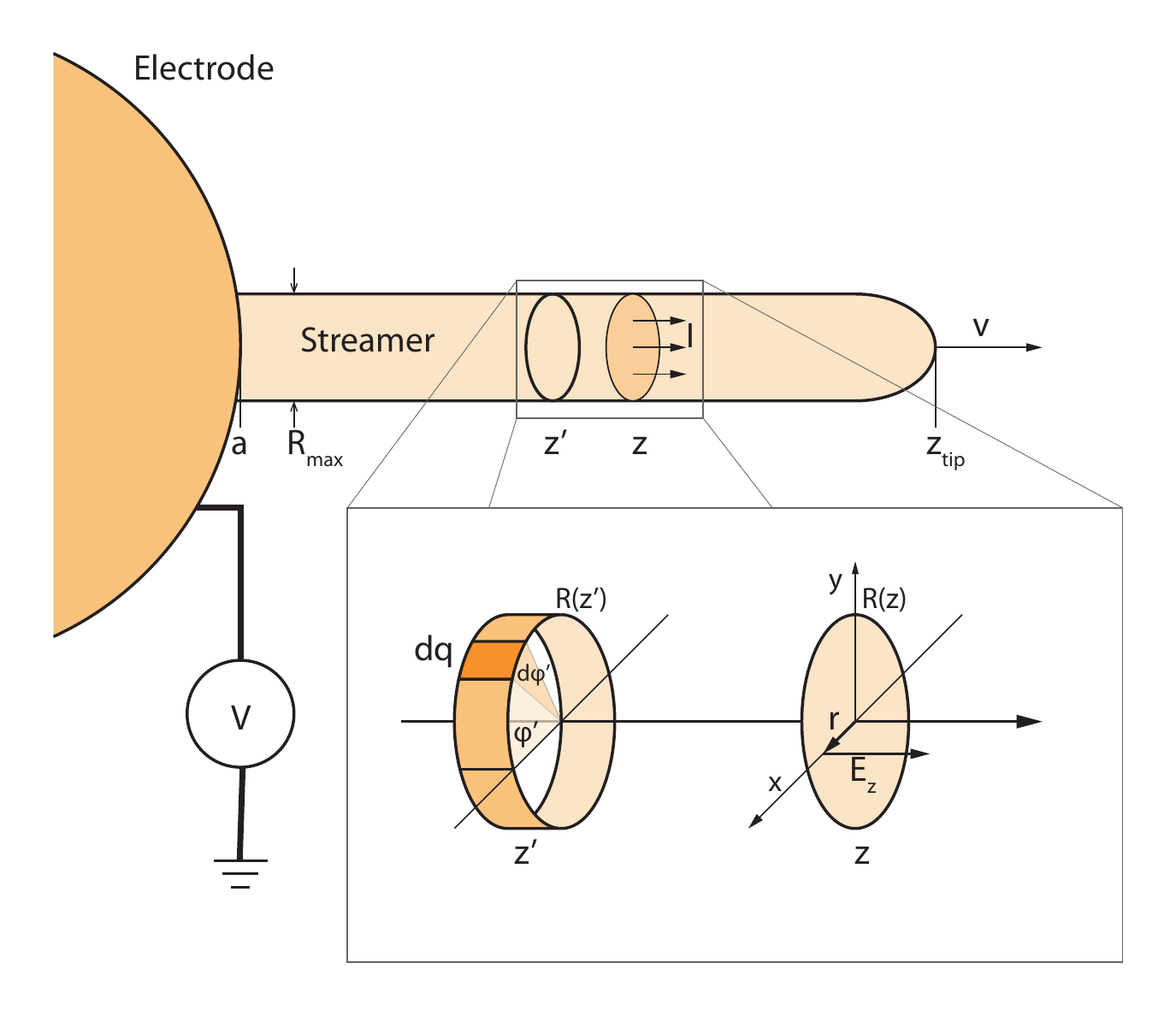}

\caption{\label{fig:Schema}
  Schematic picture of our streamer model.  We simulate a streamer that emerges
from spherical electrode at a prescribed electrostatic potential
$V$.  The electrode has a radius $a$ and is centered at the origin.  
The streamer advances with velocity $v$ in the $z$-direction and its
tip is located at the time-dependent position $z_\text{tip}$.  Far to the left
of the tip, the streamer channel assymptotocally approaches a maximum radius
$R_\text{max}$.  The inset shows the geometry the of electrostatic interaction
whereby a charge element $\dif q$ at $z'$ contributes to 
the electric field at $z$ and thus to the electric current $I$ at that point.
}
\end{figure}

Figure~\ref{fig:Schema} shows a 
schematic view of our model.  Although our approach can be generalized to other
contexts, we focus on streamers in air at atmospheric pressure.
We model the streamer as an axially symmetrical filament that grows 
in the $z$ direction due to the electric field created by a spherical 
electrode to which it is connected.  At a given time $t$ the streamer 
spans the distance from the electrode boundary $a$ to the location of the 
streamer tip $z_\text{tip}$ and propagates at a velocity
\begin{linenomath}
\begin{equation}
  v = \od{z_\text{tip}}{t}.
  \label{velocity}
\end{equation}
\end{linenomath}

The streamer shape is defined by its radius $R(z)$ in the 
range $a<z<z_\text{tip}$.  
In the simplest case we prescribe $R(z)$ to have a smooth shape around 
$z_\text{tip}$ and asymptotically approach a prescribed function
$R^\star(z)$ far from the tip.  A simple expression with these 
properties is
\begin{linenomath}
\begin{equation}
  R(z) = R^\star(z) \left(1-e^{(z-z_\text{tip})/R^\star(z)}\right)^{1/2}.
  \label{R(z)}
\end{equation}
\end{linenomath}
At the streamer tip this shape yields a radius of curvature 
$R^\star(z_\text{tip})/2$ so $R^\star(z)$ encapsulates the evolution of the
streamer radius.  As mentioned in ref.~\citep{Luque2012/JCoPh}, finding a
physically motivated evolution for the streamer radius remains an unsolved
problem of streamer physics.  Here we will mostly impose a constant 
$R^\star(z) = R_\text{max}$, with the exception of section~\ref{sect:micro} 
where, to
properly compare with a microscopic simulation, we impose that the radius grows
at a constant rate in space, $R^\star(z) = R_0 + K z$, where $K$ is
obtained from the microscopic simulation.

Our key assumption is that the streamer is so thin that we can consider
that charge transport in the transversal direction occurs instantaneously.  
In that case all the 
electric charge accumulates at the streamer's boundaries.  This behaviour is
observed in all microscopic streamer simulations (e.g. 
refs.~\citep{Dhali1985/PhRvA,Pancheshnyi2003/JPhD,Montijn2006/JCoPh,
Luque2007/ApPhL,Babaeva2009/PSST/1,Luque2010/GeoRL}).  Under this assumption the full electrodynamic state of the 
streamer can be described by a linear charge density $\lambda$ that 
satisfies
\begin{linenomath}
\begin{equation}
  \pd{\lambda}{t} = -\pd{I}{z},
  \label{lambda}
\end{equation}
\end{linenomath}
where $I$ is the electric current flowing through the streamer 
cross section.  As we discuss below, the current $I$ must include not only
the volume current flowing through the streamer body but also a surface 
current located at the streamer boundary.  We call these two components, 
respectively, \textit{channel current}, $I_C$, and \textit{surface current},
$I_S$.

\subsubsection{The channel current.}
\label{sect:channel}
This current is related to the electric current density $\mathbf{j}$ by 
an integral
over the channel's cross-section:
\begin{linenomath}
\begin{equation}
  I_C = \int_0^{R(z)} j_z \, 2\pi r \dif r.
  \label{Itotal}
\end{equation}
\end{linenomath}
The current density $\mathbf{j}$
results from drift and diffusion of all charged species $s$ within the streamer:
\begin{linenomath}
\begin{equation}
  \mathbf{j} = \sum_{s} \left(|q_s| \mu_s n_s \mathbf{E} - \nabla \cdot D_s n_s\right),
\label{full_current}
\end{equation}
\end{linenomath}
where $\mathbf{E}$ is the local electric field and 
$q_s$, $\mu_s$, $n_s$ and $D_s$ are respectively the charge,
mobility, density and diffusion coefficient of species $s$.  

To obtain a model
that can be simulated efficiently and is expected to scale to multi-streamer 
simulations, we introduced a number of simplications.  First, we neglect
diffusion 
\footnote{The relative importance of advection versus diffusion is measured
by the P\`eclet number $\mathrm{Pe}=Lu/D$, where $L$ and $u$ are, respectively 
the characteristic length and velocity of the problem and $D$ is the diffusion 
coefficient.  In our case we have $u\approx\SI{e5}{m/s}$, 
$D\approx\SI{0.2}{m^2/s}$ \cite{Dujko2011/JaJAP} so diffusion is only relevant when
$\mathrm{Pe}\lesssim 1$, at length scales smaller than about \SI{2}{\micro m}.}
Also, as the 
inner electric field within a streamer does not exhibit too large a 
variability, almost always ranging from \SI{3}{kV/cm} to 
\SI{30}{kV/cm}, we approximate the mobility of all species $\mu_s$ to be
independent of the electric field.  This assumption turns 
(\ref{lambda}) into a linear differential equation, heavily simplifying 
its solution.  A final simplification that we take for the sake of computing 
efficiency is that the species densities $n_s$ are uniform across the
channel and can be taken out of the integral (\ref{Itotal}).
Below we show that this yields a closed-form expression for one 
integral in a multi-dimensional integral expression, saving us one numerical 
integration.

With these simplifications (\ref{Itotal}) reads
\begin{linenomath}
  \begin{equation}
    I_C(z) = \sigma(z) \int_0^{R(z)} E_z(z, r) \, 2\pi r \dif r,
    \label{Isimple}
  \end{equation}
\end{linenomath}
where
\begin{linenomath}
  \begin{equation}
    \sigma(z) = \sum_s q_s \mu_s n_s(z) 
    \label{sigma}
  \end{equation}
\end{linenomath}
is the channel conductivity.

We calculate the electric field in (\ref{Isimple}) by 
decomposing it as $E_z = E_{0z} + E_{1z}$, where $E_0$ is the
background field and $E_1$ is the self-consistent field created
by the charges in the channel.  The linearity of (\ref{Isimple}) translates this
decomposition into a current $I_{C0}$ driven by the external field and a current
$I_{C1}$ due to interactions between channel elements.

For the moment, we leave aside the current driven by the external field, 
$I_{C0}$, which we more conveniently discuss in 
section~\ref{sect:electrode}, after we have also discussed the surface
current in~\ref{sect:surface}.  

Focusing on the channel current $I_{C1}$, which depends on the 
self-consistent field $E_1$, we consider the geometry in the inset of 
figure~\ref{fig:Schema}, where we are interested in the electric field at 
longitudinal coordinate $z$ and at distance $r$ away from the axis.  
To calculate this, we integrate the contributions of all 
charge elements $\dif q$ 
at longitudinal coordinates $z'$.  The charge in $\dif q$ is
\begin{linenomath}
  \begin{equation}
    \dif q = \frac{1}{2 \pi} \lambda \dif \varphi' \dif z',
    \label{dq}
  \end{equation}
\end{linenomath}
where $\varphi'$ is the azimuthal angle of the charge element.  Let us first
focus on electrostatic interactions in free space (i.e. in the absence of 
any electrode): the presence of an electrode is
discussed in the following section.  In free space the contribution of 
$\dif q$ at $z'$ to $E_{1z}$ at $z$ reads
\begin{linenomath}
  \begin{equation}
    \dif E_{1z} =  \frac{(z-z')\dif q}{4\pi \varepsilon_0\left[(r - R(z')\cos\varphi')^2 + R(z')^2\sin^2\varphi' + (z-z')^2 \right]^{3/2}}.
    \label{dEz0}
  \end{equation}
\end{linenomath}
In order to simplify our notation, it is convenient to introduce
$x(z', \varphi') = R(z')\cos\varphi'$, 
$\rho(z', \varphi') = \left(R(z')^2\sin^2\varphi' + (z-z')^2\right)^{1/2}$.  For brevity we leave 
the dependence
on $z'$ and $\varphi'$ implicit and write simply $x$ and $\rho$.  With this
notation and combining (\ref{dq}) and (\ref{dEz0}) into (\ref{Isimple}) we obtain
\begin{linenomath}
  \begin{equation}
    I_{C1}(z) = \frac{\sigma(z)}{4\pi \varepsilon_0} \int_0^{R(z)} \dif r 
               \int_{z_b}^{z_\text{tip}} \dif z' 
               \int_{0}^{2\pi} \dif \varphi' 
               \frac{(z-z')\lambda(z')r}{\left[(r - x)^2 + \rho^2 \right]^{3/2}}.
    \label{Ifull}
  \end{equation}
\end{linenomath}

As we mentioned above, one of the three integrals in (\ref{Ifull}) can be solved
analytically into a closed-form expression.  For this we make use of the 
indefinite integral
\begin{linenomath}
  \begin{equation}
    \int \frac{r\dif r}{\left[(r - x)^2 + \rho^2\right]^{3/2}} 
      = \frac{x (r - x) - \rho^2}{\rho^2 \sqrt{\rho^2 + (r - x)^2}} + C
    \label{indefint}
  \end{equation}
\end{linenomath}
and rewrite (\ref{Ifull}) as
\begin{linenomath}
  \begin{equation}
    I_{C1}(z) = \left.\frac{\sigma(z)}{4\pi\varepsilon_0} \int_{z_b}^{z_\text{tip}} \dif z' 
               (z-z')\lambda(z')
               \int_{0}^{2\pi} \dif \varphi' 
               \frac{x (r - x) - \rho^2}{\rho^2 \sqrt{\rho^2 + (r - x)^2}}
                 \right|_{r=0}^{r=R(z)}.
    \label{IF}
  \end{equation}
\end{linenomath}
Thus, defining a kernel
\begin{linenomath}
  \begin{equation}
    G_C(z, z') = (z-z') \int_{0}^{2\pi} \dif \varphi' \left.\frac{x (r - x) - \rho^2}{\rho^2 \sqrt{\rho^2 + (r - x)^2}}
                 \right|_{r=0}^{r=R(z)},
    \label{GC}
  \end{equation}
\end{linenomath}
we write (\ref{IF}) as
\begin{linenomath}
  \begin{equation}
    I_{C1}(z) = \frac{\sigma(z)}{4\pi\varepsilon_0} \int_{z_b}^{z_\text{tip}} \dif z'
      G_C(z, z') \lambda(z').
    \label{IC1}
  \end{equation}
\end{linenomath}

\begin{figure}
  \includegraphics[width=0.9\columnwidth]{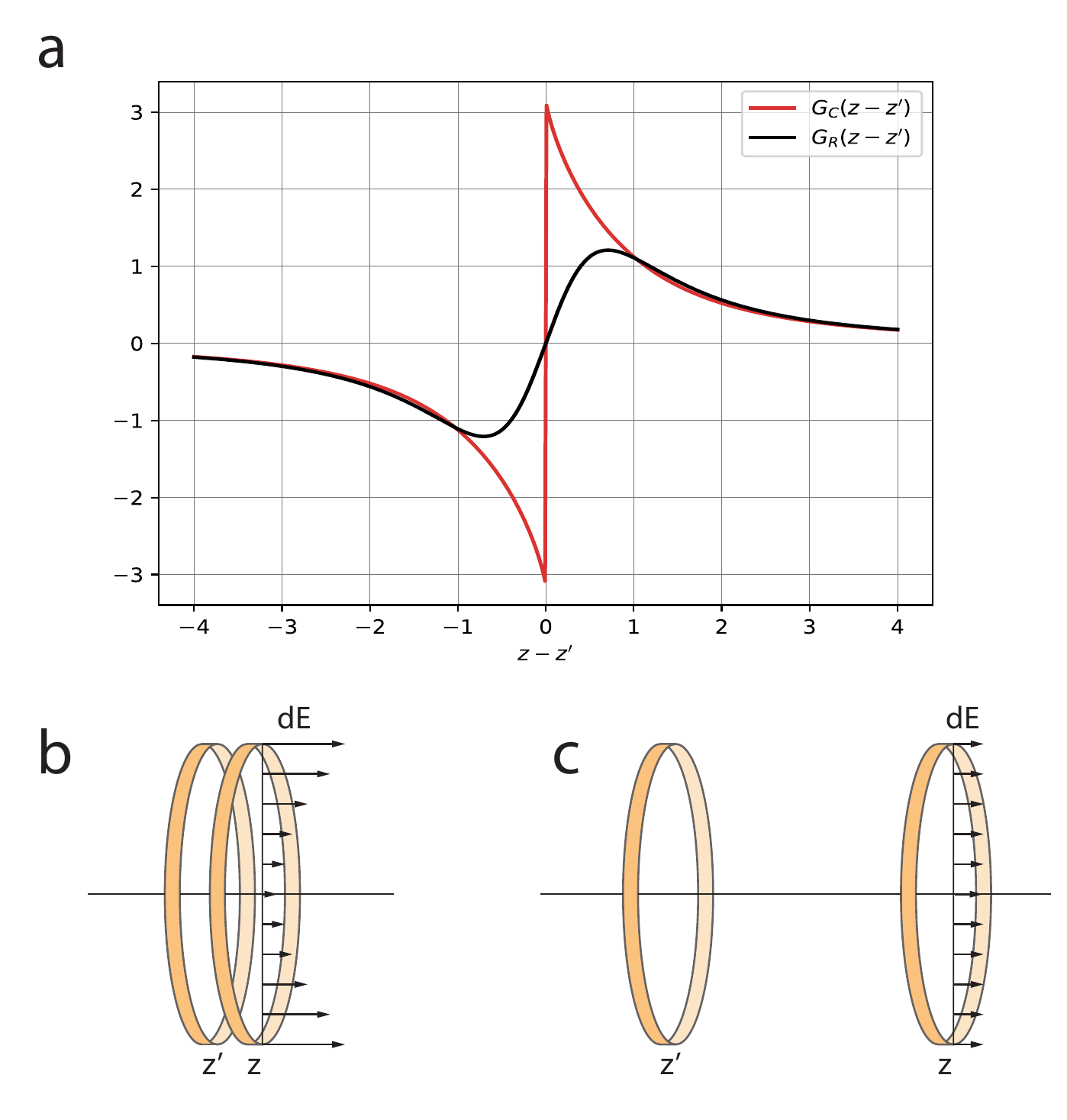}
  \caption{\label{fig:Interactions_kernels}
    (a) Interaction kernels $G_C$ and $G_R$ defined by 
    (\ref{GC}) and (\ref{G_R}) in the text, plotted here for $R(z)=1$.
    (b) The short-range interaction in the streamer channel
    is dominated by electric fields and conduction currents off-axis; 
    the ring kernel $G_R$ underestimates this interaction.  (c)
    When the two annular sections are far appart the interaction field
    is transversally homogeneous and both kernels $G_C$ and $G_R$ give similar
    results.
  }
\end{figure}

Some comments about this electrodyamic model are in order:
\begin{enumerate}
\item The integrand in (\ref{IF}) diverges as $z'\to z$ and $\varphi' \to 0$.  This
  of course stems from the divergence of the electric field around a point 
  charge.  However, one can prove that this divergence is integrable and
  the expressions (\ref{IF}) and (\ref{IC1}) are well defined.  Here we are
  calculating the field close to a surface with a smooth charge density,
  which is finite and well defined.
\item Microscopical simulations of streamers show that the electric field
  inside the streamer channel is transversally quite homogeneous.  
  One is therefore
  tempted to skip the integrals in $r$ and $\varphi'$ and take the electric 
  field at 
  the central axis as a good approximation.  This approach, called
  \emph{ring method} was employed e.g. by ref.~\citep{Aleksandrov1996/JPhD} and 
  is equivalent to replacing $G_C(z, z')$ in (\ref{IC1}) by 
  \begin{linenomath}
    \begin{equation}
      G_R(z, z') = \frac{\pi R(z)^2(z-z')}{\left[R(z')^2 + (z-z')^2\right]^{3/2}}.
      \label{G_R}
    \end{equation}
  \end{linenomath}
  However,
  as mentioned in ref.~\cite{Luque2014/NJPh}, this approximation often leads
  to unrealistic oscillations in the presence of strong longitudinal 
  inhomogeneities such as the streamer head itself.
  A comparison between $G_C(z, z')$ and $G_R(z, z')$, as shown in 
  figure~\ref{fig:Interactions_kernels}a, hints at an explanation.
  In the figure, where we have set $R(z)=1$ so that
  $G_C$ and $G_R$ become functions only of $z'-z$, we see that the kernel
  $G_R$ vanishes as $z'\to z$, which means that it neglects interactions
  between closely spaced rings in the streamer channel.  As pictured in
  figure~\ref{fig:Interactions_kernels}b, these interactions are dominated by the
  electric field away from the central axis; only when $z'-z \gg R$ can we 
  take (figure~\ref{fig:Interactions_kernels}c) the electric field in the axis as
  representative of the full cross-sectional interaction.

  Our kernel $G_C$, defined by (\ref{GC}), is discontinuous and correctly accounts
  for interactions between neighboring points.  This is necesary to 
  dynamically remove unphysical oscillations with wavelengths of the order of
  the streamer radius $R$.
  
\end{enumerate}

\subsubsection{The surface current.}
\label{sect:surface}
\begin{figure}
  \includegraphics[width=\columnwidth]{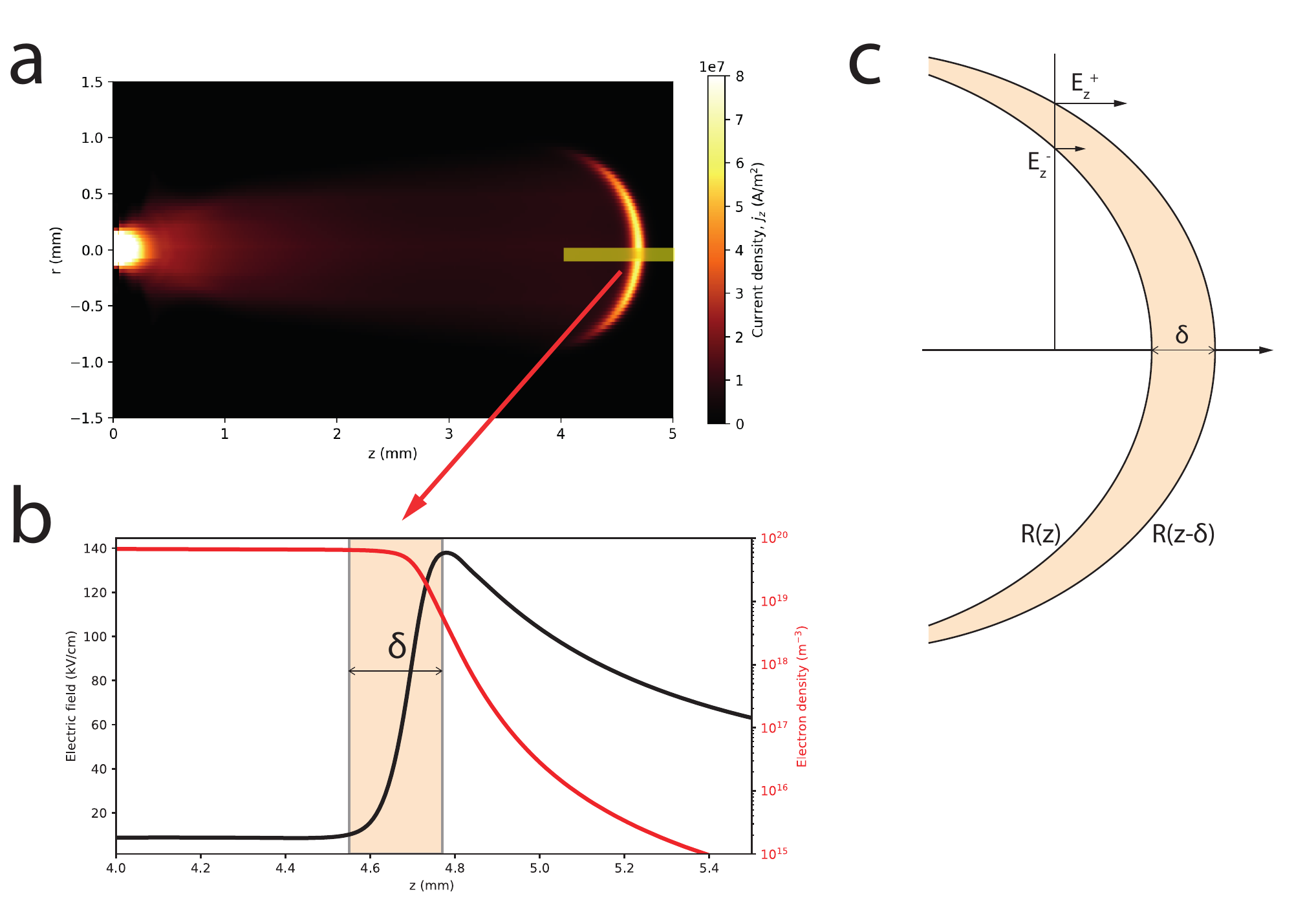}
  \caption{\label{fig:layer_current_full}
    (a) A microscopic streamer simulation shows that close to the streamer tip
    there is a layer of electric current concentrated in a thin layer at the 
    boundary.  Details for this microscopic simulation are provided in section
    \ref{sect:micro}. (b) Profiles of the electric field and electron density
    on the central axis in the microscopic simulation.  The approximate 
    width of the current sheet ($\delta$) is indicated by the shaded region.
    (c) Scheme for the integration scheme of the current sheet used in our
    macroscopic model.
  }
\end{figure}

Besides the channel current described above, a streamer also contains a sheet of
current around its head.  This current, which we name here surface current, 
is apparent in 
figure~\ref{fig:layer_current_full}a, where we show the electric current density
obtained in a microscopic streamer simulation.  The surface current is the 
main responsible of moving the space charge layer forward and it results from
the continuous growth of the streamer channel.  
Figure~\ref{fig:layer_current_full}b provides a microscopical interpretation
of the suface current: the electric field is not fully screened close to the
streamer head but rather penetrates a width $\delta$.
Within
this distance the electron density is already much higher that in the background
so the penetration of the electric field results in a significant current.

To incorporate the surface current in our one-dimensional model we need first to
estimate the width $\delta$ and then, in order to integrate across the channel,
introduce reasonable assumptions about the electric field and electron 
density within the layer.

To estimate $\delta$ we note that the penetration of the field is a consequence
of the finite conductivity of the channel combined with the streamer velocity.
If we assume that within the streamer the field follows a dielectric 
relaxation with a characteristic time $\tau=\epsilon_0 / \sigma(z_\text{tip})$,
the width of the current layer is $\delta = \xi v \tau$, where $v$
is the streamer velocity and $\xi$ is a parameter of order unity that corrects
for the curvature of the streamer head and the fact that the conductivity is
not constant along the layer's width.  In our microscopic tests we found
$\xi \approx 4$.

Figure~\ref{fig:layer_current_full}c illustrates the transversal integration
of the surface current at a given $z$.  We approximate
the channel conductivity ($\sigma$) and the $z$-component of the electric 
field ($E_z$) as linear functions between the inner and outer radius of the layer,
respectively $R^-$ and $R^+$:
\begin{linenomath}
\begin{equation}
  E_z(r) = E_z^- - \frac{(E_z^- - E_z^+)(r - R^-)}{R^+- R^-},
  \label{E_layer}
\end{equation}
\begin{equation}
  \sigma(r) = \sigma^- \frac{R^+ - r}{R^+ - R^-},
  \label{r_layer}
\end{equation}
\end{linenomath}
where $E^-$ and $E^+$ are the inner and outer values of the $z$-component 
of the electric field and where $\sigma^-$ is the inner conductivity, the outer
conductivity being neglected.

We can apply (\ref{E_layer}) and (\ref{r_layer}) to integrate the 
electric current density $j_z = \sigma E_z$ across the channel width:
\begin{linenomath}
  \begin{equation}
    I_S = 2 \pi \int_{R^-}^{R^+} E_z(r) \sigma(r) r\, \dif r = 
        \frac{\pi}{6} \sigma^- (R^+ - R^-) \left[E^+ (R^+ - R^-)
                                                + E^-(3R^- + R^+)\right].
    \label{IS}
  \end{equation}
\end{linenomath}

We incorporate (\ref{IS}) into our model by setting $R^- = R(z)$, 
$R^+ = R(z-\delta)$, $\sigma^- = \sigma(z)$ and using (\ref{dEz0}) evaluated
at $r=R(z) - \epsilon$ for $E^-$ and $r = R(z) + \epsilon$ for $E^+$,
where $\epsilon$ is a small length that captures the discontinuity
in the electric field at both sides of the thin charged layer
\footnote{Another option would be to use $R^+$ and $R^-$ also for the 
evaluation of the electric field but we note that
in our model the space charge is concentrated within an infinitely thin
layer around the streamer so we feel that using the jump of electric field
better follows the spirit of the model.  In any case since $\delta$ is small
compared with our typical distances both approaches produce very similar
results.}.  We take $\epsilon = \SI{10}{\micro m}$. 

Finally, we cast expression (\ref{IS}) into the same form as (\ref{IC1}) by
noting that (\ref{IS}) is linear in $E^+$ and $E^-$.   We find
\begin{linenomath}
  \begin{equation}
    I_{S}(z) = \frac{\sigma(z)}{4\pi\varepsilon_0} \int_{z_b}^{z_\text{tip}} \dif z'
      G_S(z, z') \lambda(z').    
    \label{ISG}
  \end{equation}
\end{linenomath}
where
\begin{linenomath}
  \begin{equation}
    G_S(z, z') = \frac{\pi}{6} (R^+ - R^-) \left[U^+(z, z') (R^+ - R^-)
      + U^-(z, z')(3R^- + R^+)\right],
    \label{GS}
  \end{equation}
\end{linenomath}
\begin{linenomath}
  \begin{equation}
    U^\pm(z, z') = (z-z') \int_{0}^{2\pi} \dif \varphi' 
      \frac{1}{2\pi \left[(R(z) \pm \epsilon - x)^2 + \rho^2 \right]^{3/2}}.
  \end{equation}
\end{linenomath}
Combining expressions (\ref{IC1}) and (\ref{ISG}) we calculate the total
self-consistent current from a single kernel 
$G(z, z') = G_C(z, z') + G_S(z, z')$:
\begin{linenomath}
  \begin{equation}
    I_{C1} + I_S = \frac{\sigma(z)}{4\pi\varepsilon_0} \int_{z_b}^{z_\text{tip}} \dif z'
      G(z, z') \lambda(z').
    \label{IG}
  \end{equation}
\end{linenomath}

\subsubsection{Background field and inclusion of an electrode.}
\label{sect:electrode}
In most experiments, streamers start from an enhanced electric field
around a high-voltage, pointed electrode.  To reproduce this setup we consider
here that the streamer emerges from a spherical electrode at an electrostatic 
potential $V$ (see figure~\ref{fig:Schema}) that is centered at the 
origin and has a radius $a$.  In our model, we account for 
this electrode in two places: (a) in the background electric field $E_0$ 
introduced earlier and (b) in a modification of the kernel in 
(\ref{IG}) to include the effect of mirror charges required to satisfy the boundary conditions imposed
by the electrode.

For the first point (a), the component of the total current due to 
the background field is what we called $I_{C0}$ in section~\ref{sect:channel}.
It can be calculated by integrating the $z$-component of 
the electric field created by the electrode, which yields
\begin{linenomath}
  \begin{equation}
    I_{C0}(z) = 2\pi\sigma(z) a V \left(1 
      - \frac{z}{\left(R(z)^2 + z^2\right)^{1/2}}\right).
    \label{IC0}
  \end{equation}
\end{linenomath}
\begin{figure}
  \includegraphics[width=0.9\columnwidth]{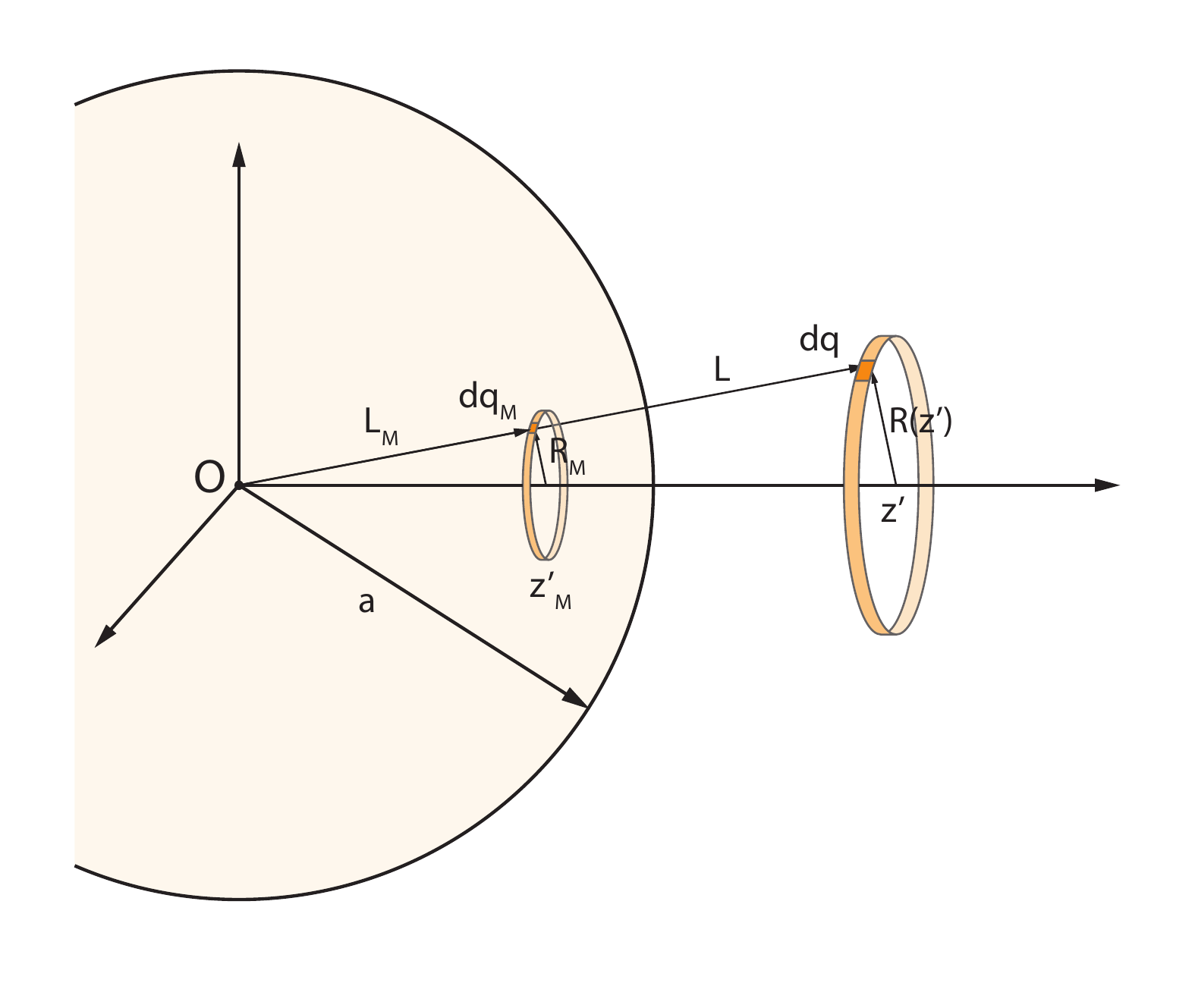}
  \caption{\label{fig:Mirror}
    Computation of mirror charges required to satisfy the boundary conditions
    of an electrode of radius $a$.  Here we consider a charge element $\dif q$
    at $z'$, where the channel radius is $R(z')$.  The boundary condition
    imposed by the presence of a spherical electrode located at the origin
    are satisfied by including a mirror charge $\dif q_M$ as described in the
    text.
  }
\end{figure}

Turning now to (b), in order to calculate the effect of mirror charges 
we consider 
the geometry shown in figure~\ref{fig:Mirror}, where a charge element 
$\dif q$ sits at axial coordinate $z'$ and radius $R(z')$.  The boundary
condition imposed by the presence of the electrode is satisfied if we include
a mirror charge $\dif q_M$ located on the line that joins the electrode's center
and $\dif q$ and at a distance $L_M$.  Following e.g. ref.~\citep{Jackson1975/book}
we find that $\dif q_M = -\kappa \dif q$, $L_M = \kappa^2 L$, where 
$L = \left(z'^2 + R(z')^2\right)^{1/2}$ and $\kappa = a / L$.  The $z$-coordinate
 of $\dif q_M$ is thus $z'_M = \kappa^2 z'$.  Therefore we include the effect of 
the electrode if we 
update the kernel function $G$ in (\ref{IG}) as
\begin{linenomath}
  \begin{equation}
    \bar{G}(z, z') = G(z, z'; R(z')) - \kappa G(z, \kappa^2 z'; \kappa^2 R(z')),
    \label{Gbar}
  \end{equation}
\end{linenomath}
where we made the dependence on $R(z')$ explicit.  Henceforth we calculate the self consistent current using $\bar{G}$ instead of $G$ in equation (\ref{IG}).

\subsection{Chemical processes and mobilities}
\label{sect:propagation}

In general, many chemical reactions between active species operate 
within the streamer channel.  These reactions influence the channel 
conductivity and must therefore
be coupled to the electrodynamic evolution described in the previous section.
Here we considered a chemical model composed of 
17 species coupled through 78 reactions detailed in the supplementary material.
The chemical model focuses on the evolution of electron density and ionic 
species following references 
\cite{Kossyi1992/PSST,Aleksandrov1999/PSST,Pancheshnyi2013/JPhD} and includes
the effect of water vapor as modeled by Gallimberti 
\cite{Gallimberti1979/JPhys}.
 Note that this chemical model is designed to investigate changes in the conductivity for longer timescales than those considered in this work and thus many of the included reaction play a negligible role.  Nevertheless we opted for keeping
them as a reference.  

The chemical model determines the evolution of the density of each species $s$
as
\begin{linenomath}
  \begin{equation}
    \pd{n_s}{t} = C_s = \sum_{r \in \text{reactions}}A_{sr} k_r n_{\mathcal{I}(r,1)} n_{\mathcal{I}(r,2)}\dots,
    \label{reactions}
  \end{equation}
\end{linenomath}
where $C_s$ is the net creation of species $s$, $A_{sr}$ is the net number
of molecules of species $s$ created each time that reaction $r$ takes place,
$k_r$ is the rate coefficient of reaction $r$ and 
$\mathcal{I}(r,1), \mathcal{I}(r,2), \dots$ are the indices of the input species
of reaction $r$.  Here the rate coefficient $k_r$ is, in general, a 
function of the local electric field.  Since the transversal variation
of the electric field is dynamically suppresed by the kernel described in 
the previous section, here it is justified to calculate the rate coefficients
from the electric field at the streamer axis.  Thus $k_r$ depends on
\begin{linenomath}
  \begin{equation}
    E_\text{axis}(z) = \frac{1}{4\pi\varepsilon_0} \int_{z_b}^{z_\text{tip}} \dif z'
      \bar{G}_R(z, z') \lambda(z'),
    \label{Eaxis}
  \end{equation}
\end{linenomath}
where $\bar{G}_R(z, z')$ is the kernel function obtained from (\ref{G_R}) by
adding the effect of mirror charges as in (\ref{Gbar}).

As the streamer propagates (see next section), it changes the composition of
the gas ahead of its tip through photo-ionization and the enhancement 
of the electric field.  Our model does not include the
dynamics ahead of the streamer tip so the effect of these processes is
modeleled by imposing densities $n_s^0$ for each species $s$ at the streamer tip 
$z_\text{tip}$.  We consider that the pre-streamer dynamics elevate the
electron density to a prescribed value $n_e^0$; to ensure quasi-neutrality 
this density is balanced by concentrations of
\ce{O2+} and \ce{N2+}
that follow the relative densities of 
\ce{O2} and \ce{N2} in air.  The densities of all other species are set to zero
at $z_\text{tip}$.

All charged species contribute to the channel conductivity, which we calculate
with (\ref{sigma}).  We take the electron mobility as 
$\mu_e = \SI{380}{cm^2/V/m}$ \cite{Dhali1987/JAP}.  For \ce{O-}, \ce{O2-} and
\ce{O3-} we use values from ref.~\cite{Viehland1995/ADNDT} fetched 
from the LxCat
database \cite{Pancheshnyi2012/CP} selecting the approximate mobilities 
for a reduced electric field of \SI{100}{Td}.  This gives us
\begin{linenomath}
  \begin{align}
    \mu_{\cee{O-}} &= \SI{4.5}{cm^2/V/m}, \nonumber \\
    \mu_{\cee{O2-}} &= \SI{2.7}{cm^2/V/m}, \nonumber \\
    \mu_{\cee{O3-}} &= \SI{2.8}{cm^2/V/m}.
    \label{mobilities}
  \end{align}
\end{linenomath}
Within our model's accuracy, all other ions, including water cluster ions
\cite{Wissdorf2013/JASMS}, can be assumed to have roughly the same mobility,
which we take as
\begin{linenomath}
  \begin{equation}
    \mu_{\text{ion}} = \SI{2}{cm^2/V/m}.
    \label{mobility_ions}
  \end{equation}
\end{linenomath}

\subsection{Streamer propagation}
\label{sect:propagation}
At the same time that charge is transported and chemical reactions are operating
within the streamer channel, the streamer tip advances.  
It is generally accepted that the speed of this advance, as defined in 
(\ref{velocity}), depends on the streamer's radius and the electric field at 
its tip, $E_\text{tip}$.  This is,
\begin{linenomath}
  \begin{equation}
    \od{z_\text{tip}}{t} = v(E_\text{tip}, R_\text{max}).
    \label{dztip}
  \end{equation}
\end{linenomath}
Here $E_\text{tip}$ can be evaluated 
from (\ref{Eaxis}) as $E_\text{tip} = E_\text{axis}(z_\text{tip}^+)$, where 
$z_\text{tip}^+$ means that, since the field is discontinuous at $z_\text{tip}$, 
we take the value immediately outside the streamer.

Naidis \cite{Naidis2009/PhRvE} investigated the relation between streamer
radius, peak electric field and velocity.   He considered the active area 
ahead of the streamer where the electric field is above the breakdown 
threshold $E_p$.  By assuming that the multiplication factor $M$ of the electron 
density within this area (or rather, its logarithm) is roughly the same 
for all streamers, Naidis derived
the following expression for the streamer velocity $v$:
\begin{linenomath}
  \begin{equation}
    \gamma R_\text{max} E_\text{tip} \int_{E_p}^{E_\text{tip}}
       \frac{\dif E \, \nu(E)}{E^2(v \pm \mu_e E)} 
       = \log M + \log \left(\frac{v\pm\mu_e E_\text{tip}}
         {v\pm\mu_e E_p}\right), 
    \label{naidis}
  \end{equation}
\end{linenomath}
where $\gamma$ is a factor of order unity that relates the spatial
decay of the electric field to the streamer radius 
(we assume $\gamma=1/2$), $\nu(E)$ is the
field-dependent temporal growth rate of electrons and $\mu_e$ is their 
mobility.  As proposed by Naidis, we take $\log M = 8$.

The streamer velocity in our model is obtained by solving for $v$ in
(\ref{naidis}), given $R_\text{max}$ and $E_\text{tip}$.  Nevertheless, 
in section \ref{sect:results} below, we investigate
the effect of the velocity on a streamer's properties by manually tuning 
the velocity for a given peak field and radius.  With that purpose, we
multiply the velocity $v$ resulting from (\ref{naidis}) by a factor $\beta$.

\section{Numerical implementation}
\label{sect:implementation}
\begin{figure}
  \includegraphics[width=0.9\columnwidth]{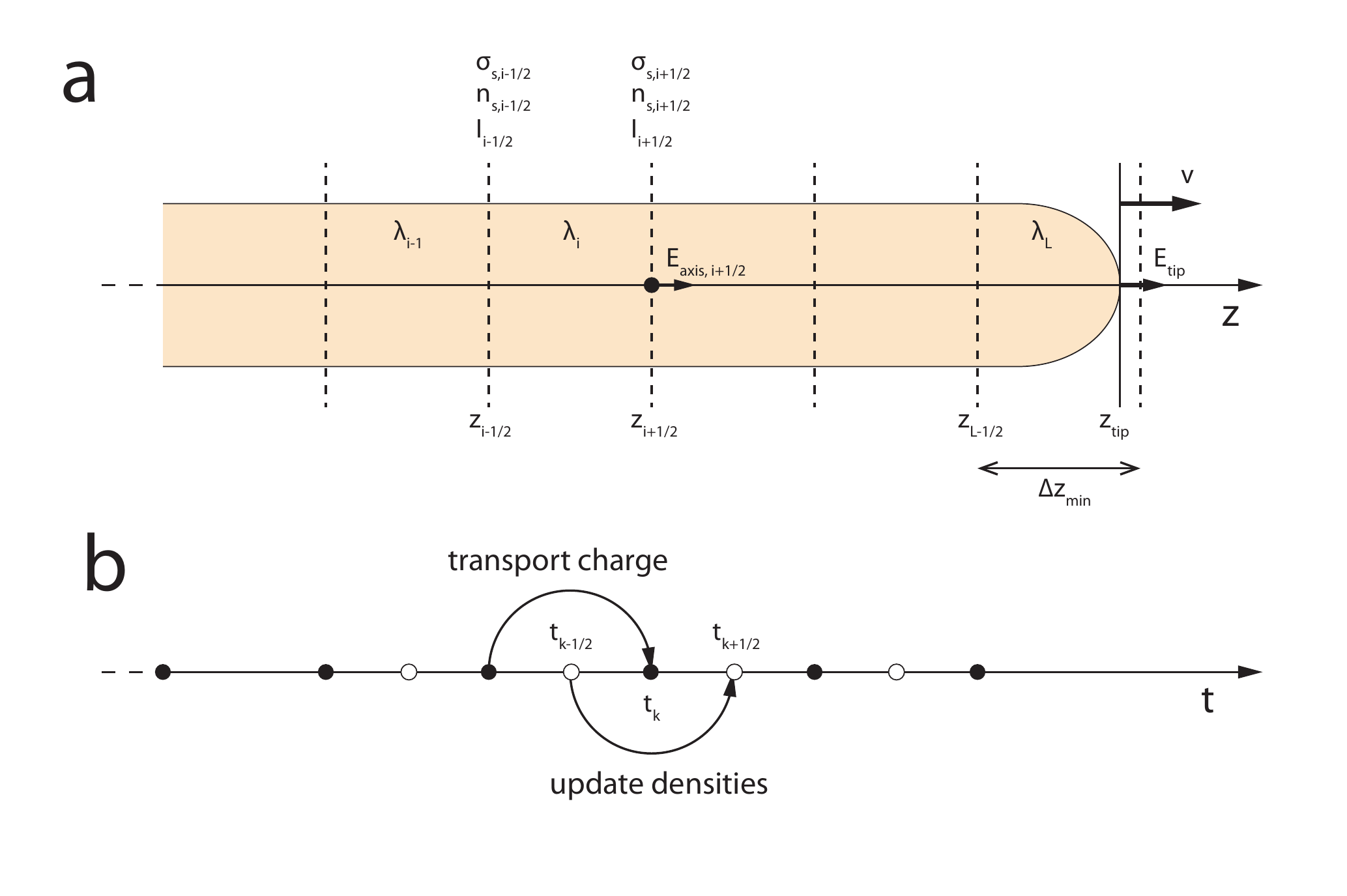}
  \caption{\label{fig:Discretization}
    Spatial (a) and temporal (b) discretization schemes for solving our model.
    As described in the text, we implemented a finite volumes method where 
    the streamer lenght is divided into $L$ cells, with average charge densities 
    defined for each cell.  At the cell boundaries we evaluate electric currents,
    species densities and conductivities.  The time integration uses a leapfrog
    scheme that alternates between solving charge transport and updating
    densities and conductivities.
  }
\end{figure}
Figure~\ref{fig:Discretization} sketches the spatial and temporal discretizations
that we implemented for the model described above.  At a given time 
the streamer length is 
divided into cells $\mathcal{C}_1, \mathcal{C}_2, \dots \mathcal{C}_L$ with
boundaries defined as $\mathcal{C}_i = (z_{i-1/2}, z_{i+1/2})\dots 
\mathcal{C}_L=(z_{L-1/2}, z_\text{tip})$.  Note that the right boundary of
the rightmost cell is $z_\text{tip}$ and that this boundary moves as the
streamer advances.  The rest of the cell boundaries are fixed within
a time step but, as described below, the mesh structure is updated at certain
times during the simulation.

To each cell we assign an average 
charge density $\lambda_i$ whereas the density of species $s$, $n_{s, i\pm 1/2}$,
and the channel 
conductivity, $\sigma_{i\pm 1/2}$, are defined at the cell 
boundaries $z_{i\pm 1/2}$.  We integrate in time using a leapfrog method, 
whereby we alternate between a step that advances the streamer head and solves 
(\ref{lambda}) for charge tansport from time $t_{j-1}$ to $t_j$ 
and a step that solves the chemical system (\ref{reactions}) 
from time $t_{j-1/2}$ to $t_{j+1/2}$.  Let us describe each of these types of
steps.

\subsection{Charge transport and streamer progression}
In the first kind of step,  we integrate the transport of charge and advance the streamer tip
from $t_{j-1}$ to $t_j$ assuming fixed particle densities and channel conductivity.  To simulate the 
transport of charge through the streamer channel we implement
a first-order accurate spatial discretization of (\ref{lambda}).
As the length of the rightmost cell of the streamer ($z_\text{tip}-z_{L-1/2}$),
changes as the tip advances, our approach is more clearly formulated 
in terms of the total charge in a cell, $q_i = \lambda_i (z_{i+1/2}-z_{i-1/2})$.  
In these
terms, the spatially discrete form of (\ref{lambda}) reads
\begin{linenomath}
  \begin{equation}
    \od{q_i}{t} = I_{i-1/2} - I_{i+1/2}.
    \label{lambda-discrete}
  \end{equation}
\end{linenomath}
In a charge-transport timestep, we integrate (\ref{lambda-discrete}) 
calculating $I_{i\pm 1/2}$ from
equations (\ref{IG}) and (\ref{IC0}).  For 
the self-consistent current 
we compute
numerically the integrals involved in (\ref{IG}) using a Gauss-Legendre 
quadrature for $z'$ within each cell and for the azimuthal 
angle $\varphi'$.  In our
first-order accurate scheme we assume a constant linear charge density inside
each cell, which leads to a linear system
\begin{linenomath}
  \begin{equation}
    I_{i+1/2} = \sum_{k=1}^{L} a_{ik}(t)q_k + b_{i}(t),
    \label{linear0}
  \end{equation}
\end{linenomath}
where the first term results from self-interaction ($I_{C1}+I_{S}$) and the second term 
from the background field ($I_{C0}$).  Even though we fix conductivities
$a_{ik}$ and $b_{i}$ change during a timestep due to the advancing
streamer tip.  Defining the matrix $\mathbf{W}(t)$ with elements
$W_{ij}(t) = a_{i-1\,k}(t) - a_{ik}(t)$ and the vector $V(t)$ with 
components $V_i(t) = b_{i-1}(t) - b_i(t)$, (\ref{lambda-discrete}) has the 
matrix form 
\begin{linenomath}
  \begin{equation}
    \od{q}{t} = \mathbf{W}(t) q + V(t).
    \label{lambda-matrix}
  \end{equation}
\end{linenomath}
This equation is coupled with equation (\ref{dztip}), which determines 
the advance of $z_\text{tip}$.  As this advance is generally smooth and not too 
far from uniform translation, an explicit Euler integration is accurate 
enough.  Since the tip velocity depends on the peak electric field
$E_\text{tip}$, we integrate (\ref{Eaxis}) at $z=z_\text{tip}^+$ also by means of
a Gauss-Legendre quadrature.

With this approach, given the status of the streamer at time $t_{j-1}$ we 
calculate $z_\text{tip}(t_j)$ and thus $\mathbf{W}(t_j)$ and $V(t_j)$.  
We then integrate (\ref{lambda-matrix}) in time with a semi-implicit 
Crank-Nicolson scheme, which yields the following linear system to obtain 
the charge at time $t_{j+1}$:
\begin{linenomath}
  \begin{equation}
    \left[\mathbf{1} - \frac{\Delta t}{2} \mathbf{W}(t_{j + 1})\right] 
    \mathbf{q}(t_{j + 1}) = 
      \left[\mathbf{1} + \frac{\Delta t}{2} \mathbf{W}(t_{j})\right] 
      \mathbf{q}(t_{j}) + \frac{\Delta t}{2} \left[V(t_{j}) + V(t_{j + 1})\right].
    \label{q-crank}
  \end{equation}
\end{linenomath}

\subsection{Update of the species densities}
Alternating with the step that we just described, we perform steps
where, for given values of $\lambda_i$ and $z_\text{tip}$ at time $t_j$, we 
update the species densities and the channel conductivity
from time $t_{j-1/2}$ to $t_{j+1/2}$.  We integrate (\ref{Eaxis}) with
a Gauss-Legendre quadrature in each spatial cell to obtain
$E_\text{axis}$ at points $z_{i\pm 1/2}$.  From this we compute
all chemical reaction rates $k_r$ in equation (\ref{reactions}).  Note that
within this kind of timestep the temporal evolution of chemical 
species at a given
point $z_{i\pm 1/2}$ is decoupled from all other points and can be solved 
independently.  Here we also apply a Crank-Nicolson scheme but in this
case this method leads to a nonlinear equation which we solve
using the Newton-Raphson method.

\subsection{Adaptation of the spatial mesh}
So far we have described the update of streamer variables within a 
fixed spatial mesh (with the exception of the right boundary
at $z_\text{tip}$). However, this scheme presents 
two problems:
\begin{enumerate}
  \item The rightmost spatial cell, bounded by $z_\text{tip}$, grows 
    disproportionally long.  To prevent this, whenever the length of this cell 
    exceeds a length $\Delta z$ we split it at 
    $z_{L-1/2} + \Delta z$ and increase the total number of cells, $L$.
    We split the total charge in the cell assuming a constant charge density
    and we interpolate linearly the values of the species densities at the newly
    created cell boundary.
  \item Generally we need a better resolution close to the streamer tip but it 
    is wasteful to use similar cell sizes along the full length of the 
    streamer.  To improve the efficiency of the code without sacrificing too much
    accuracy we employ an adaptative mesh.  Every $n_\text{coarsen}=10$ time
    steps we update our mesh by merging cells where an estimate of the 
    logarithmic slope of the absolute value of the charge density is below
    a given threshold $\epsilon_\text{coarsen}=\num{5e-2}$.
\end{enumerate}

\subsection{Implementation}
Our simulations are dominated by the computation of electrostatic interactions.
As we calculate all pairs of interactions between cells, it takes $O(L^2)$ 
computations to find the time derivative of the charge density.  Furthermore,
each of these $O(L^2)$ computations involves a two-dimensional integral
(in $z'$ and in $\varphi'$).  It is thus clear that computational efficiency
was a prime concern for us.

Fortunately most of these calculations are independent from each other and
therefore our problem is easily parallelizable.  We developed two versions
of our code: one runs in standard multicore processors and is parallelized using
OpenMP and another is implemented using the Compute Unified Device Architecture
(CUDA) and runs in General-Purpose Graphics Processing Units (GPGPUs).
As the latter version benefits from massive parallelism it runs 
between 1.5 and 14 times faster than the OpenMP version, depending on 
the resolution.

In all simulations reported here we used time-steps $\Delta t = \SI{2e-12}{s}$
and smallest spatial mesh size $\Delta z = \SI{100}{\micro m}$.

\section{Comparison with microscopic simulations}
\label{sect:micro}
\begin{figure}
\includegraphics[width=0.9\columnwidth]{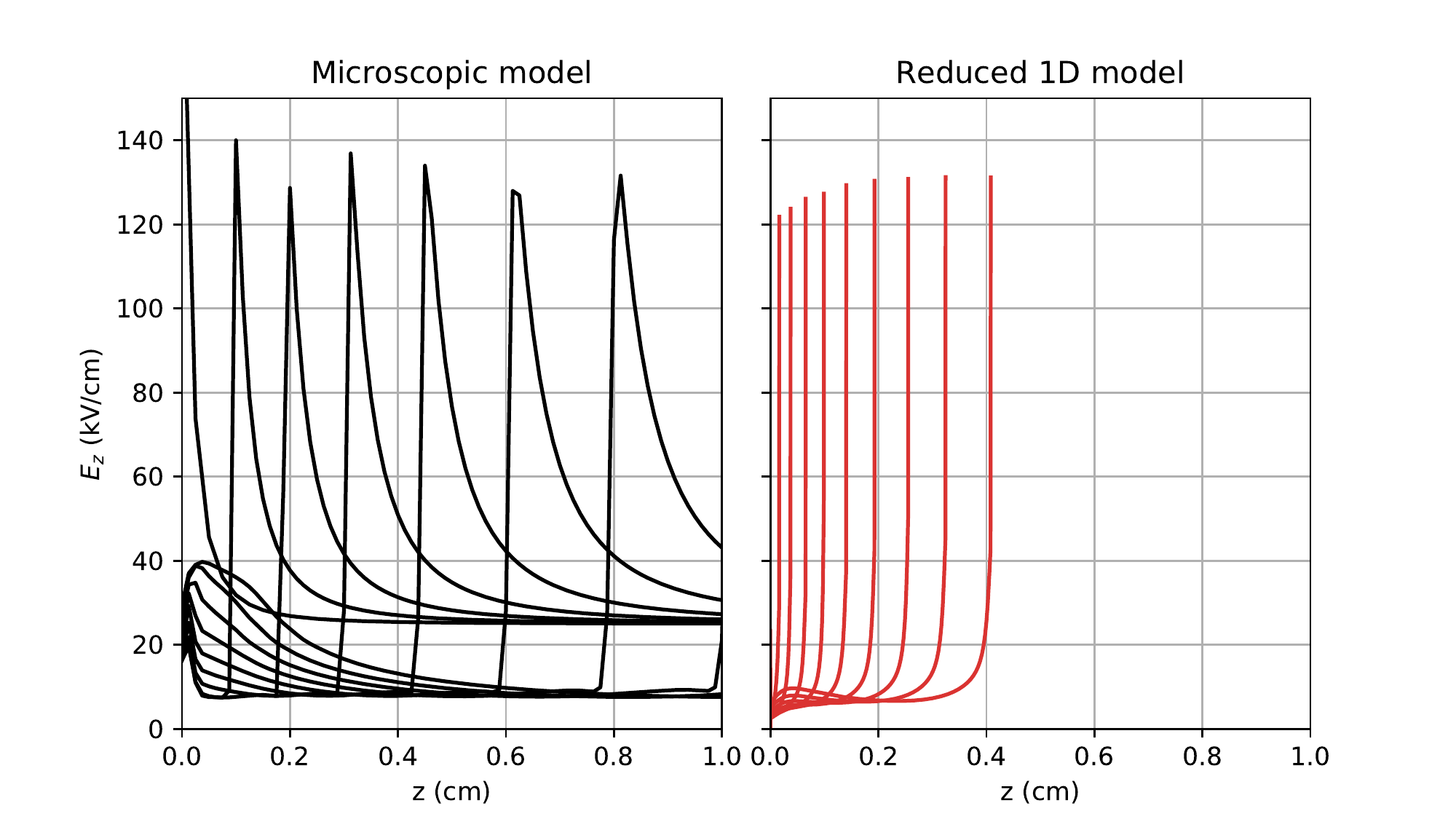}

\caption{\label{fig:compare}
  Comparison beween a microscopic streamer model (left) 
  and the 1D model presented in this work (right).  Both models are applied
  to a positive streamer propagating to the right under conditions as similar
  as possible given the differences between the two approaches.
  For each of the models we show the evolution of the axial electric field.
  All curves are plotted at regular intervals of \SI{1}{ns}.
}
\end{figure}

In this section we test the model described above and its implementation against
a microscopic streamer code.  For this purpose we use the existing
\texttt{ARCoS} code\footnote{\url{http://md-wiki.project.cwi.nl/index.php/ARCoS_code}},
which has been previously applied to problems of streamer dynamics both
at atmospheric pressure \cite{Luque2008/PhRvL,Luque2008/JPhD} and in the context
of high altitude atmospheric discharges (sprites) 
\cite{Luque2009/NatGe,Luque2010/GeoRL,Luque2011/GeoRL,Luque2016/JGRA/temp}.  
The code is based on an 
adaptive-refinement scheme \cite{Montijn2006/JCoPh} and is 
capable of working with slightly non-axisymmetric streamers and inhomogeneous 
backgrounds (for a review see ref.~\citep{Luque2012/JCoPh}).  The 
microscopic model implements a field-dependent electron mobility and includes
electron impact ionization of \ce{N2} and \ce{O2} molecules as well as
dissociative attachment to \ce{O2}.  Swarm parameters are solved offline
using Bolsig+ \cite{Hagelaar2005/PSST} with the cross-sections from 
ref.~\citep{Phelps1985/PhRvA} fetched from the LxCat database \cite{Pancheshnyi2012/CP}.

For our comparison we selected the propagation of a positive streamer at 
atmospheric pressure initiated from a needle mock-up as described in 
ref.~\citep{Luque2008/JPhD} with a needle ``lenght'' of \SI{2}{mm} and a ``radius'' of
\SI{0.2}{mm}.  We apply a potential difference of \SI{50}{kV} between this needle
and a planar electrode located \SI{2}{cm} away from the tip.  We start the
streamer placing a neutral, spherical gaussian seed with an $e$-folding 
length of \SI{0.15}{mm} and a total of \num{4.6e9} free electrons.  

Turning now to the parameters of the macroscopic, 1D model presented in this work, we simulate the
protrusion-plane geometry of the microscopic model by starting from an existing
\SI{2}{mm}-long ionized channel attached to a conducting plane, which we 
simulate by using a large electrode radius in the geometry described in 
figure~\ref{fig:Schema}.  To this configuration we apply an external
uniform background electric field of \SI{25}{kV/cm}, which coincides 
with the average electric field in the microscopic simulation.

A major problem for this comparison is that in the microscopic 
model the streamer expands significantly.  As mentioned above, 
lacking a self-consistent evolution of
streamer radius is the main limitation of our 1D model.  Nevertheless
we can check if all other features of the model are consistent with the 
microscopic simulation by externally imposing a fixed dependence of 
the tip radius 
with respect to the streamer length.  This was the motivation of introducing
$R^\star(z)$ in (\ref{R(z)}).  From the microscopic model and the 
configuration described above we found $R^\star(z)\approx R_0 + K z$ 
with $K = 0.1$ and $R_0 = \SI{0.5}{mm}$.

The results of the comparison are plotted in figure~\ref{fig:compare}, where
we show the evolution of the axial electric field and the linear charge 
density.  The figure shows that the
1D model underestimates the streamer velocity by about a factor 2.  On the other
hand, the peak electric field is very similar in both simulations.


For a given peak electric field and streamer radius the expression 
(\ref{naidis}) provides an unequivocal value for the streamer velocity.
Therefore we attribute the speed difference between the two models to
(a) inaccuracies in how we model the radius evolution in the 1D model,
in particular during the early stages of evolution and to (b) inaccuracies in 
expression (\ref{naidis}), which are likely due to the strongly nonlinear
nature of streamers and to the imprecise definition of radius in an actual 
or microscopically modelled streamer. 

\section{Results}
\label{sect:results}
As we mentioned in the introduction, one of the advantages of a simplified model
such as the one we have introduced here is that we can manually adjust
parameters that in microscopic simulations are emergent properties of the
dynamics.  This helps us to reason about the relationships between different
streamer features.

\begin{figure}
  \includegraphics[width=0.9\columnwidth]{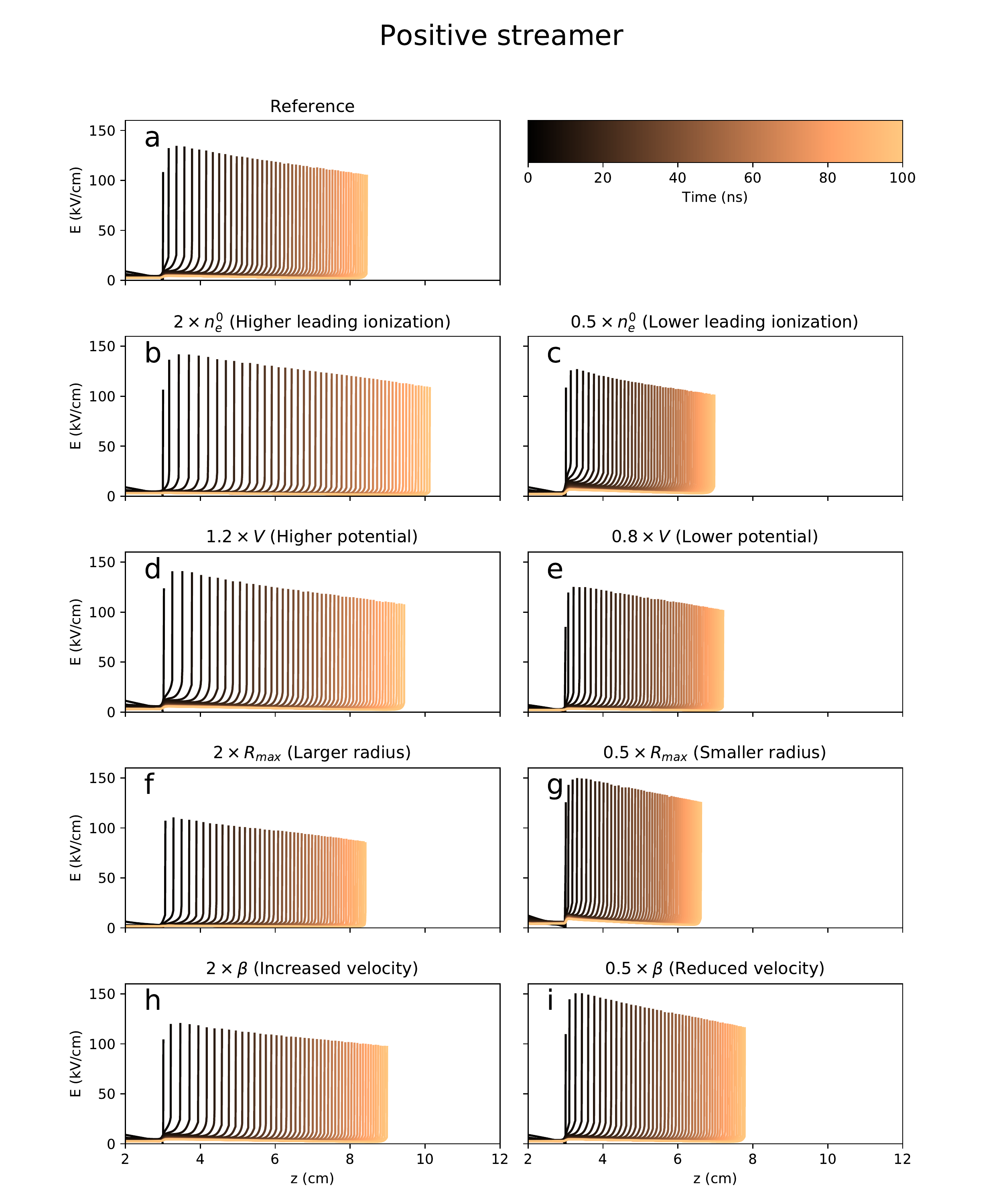}
  \caption{\label{fig:batch_positive}
    Effect of the most relevant parameters on the evolution of positive 
    streamers.  The uppermost plot shows the evolution of a streamer
    with the parameters of table~\ref{tbl:parameters}; in each subsequent
    row we have altered one of these parameters.  This change is denoted by, 
    for example, $2\times\beta$, which indicates that in the corresponding
    simulation we increased $\beta$ by a factor 2.  For each 
    configuration we show
    snapshots of the axial electric field at intervals of \SI{1}{ns}.  Each
    snapshot is colored according to its time, as indexed
    in the colorbar at the top and right.
  }
\end{figure}

\begin{figure}
  \includegraphics[width=0.9\columnwidth]{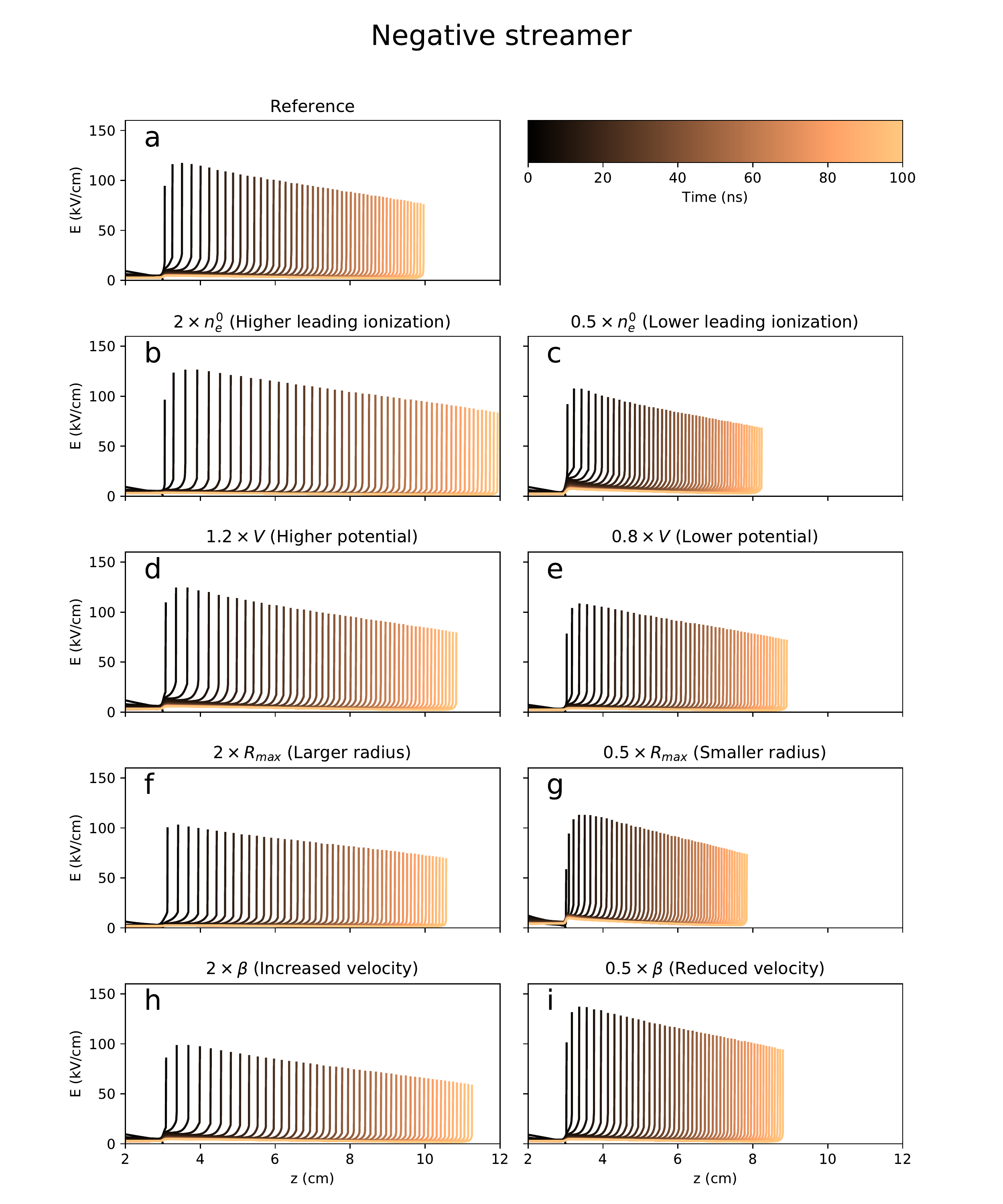}
  \caption{\label{fig:batch_negative}
    Effect of the most relevant parameters on the evolution of negative 
    streamers.  See the main text and the caption of 
    figure~\ref{fig:batch_positive}
    for a more complete description.
  }
\end{figure}

In this section we take one set of reference parameters and investigate 
how the streamer dynamics are affected by changes in the most relevant of 
these parameters.  The reference parameters are listed in 
table~\ref{tbl:parameters}.  With these parameters we run simulations
for both positive and negative streamers, the only difference between the two
being the signs in the velocity expression (\ref{naidis}).  Then we also run
simulations where we altered one of the reference parameters.  We show 
the outcome of these simulations for positive and negative streamers
in figures~\ref{fig:batch_positive} and~\ref{fig:batch_negative}.

\begin{table}
\begin{tabular}{cll}
Parameter & Description & Reference value \\ \hline
$a$ & Radius of the electrode & \SI{5}{mm} \\
$n_e^0$ & Electron density at the streamer tip & \SI{4e19}{m^{-3}} \\
$V$ & Voltage of the electrode & \SI{50}{kV} \\
$R_\text{max}$ & Largest streamer radius & \SI{1}{mm} \\
$\beta$ & Extra factor to manually change the streamer velocity & 1
\end{tabular}
\caption{\label{tbl:parameters} Main parameters for a streamer simulation in our
1D model and the values that we take as a reference to investigate their role.}
\end{table}

\subsection{Differences between positive and negative streamers}
The results in figures~\ref{fig:batch_positive} and \ref{fig:batch_negative}
show that our model is not yet predictive enough to explain some features
commonly observed in streamer experiments and simulations.  However, an
examination of these features sheds some light on streamer physics and on the
missing features of this model.

The first such feature that stands out is that in our case negative streamers
propagate faster than positive streamers, something opposite to
what is observed.  This issue is discussed e.g. in ref.~\cite{Luque2008/JPhD},
where the higher velocity of positive streamers is attributed to a higher
peak electric field, which in turns results from a sharper gradient of the 
electron density 
close to the head.  Negative streamers, where electrons
diverge from the head, have a more diffuse electron density
and thus lower electric field and slower propagation.

Microscopic simulations show that the difference in propagation 
direction of electrons relative to the streamer translates into
a higher electron density in the interior of positive streamers.  Therefore to
properly model differences between positive and negative streamers
we have to not only change the signs in (\ref{naidis}) but also
our parameter $n_e^0$, which describes the multiplication of electrons ahead
of the streamer.   

\subsection{Boundary electron density}
As shown in figures~\ref{fig:batch_positive} and \ref{fig:batch_negative},
a change in $n_e^0$, which stands for the electron density at the streamer
tip, has a significant effect on the properties of both positive and negative 
streamers.  A higher leading ionization produces a stronger field enhancement
and faster streamer propagation.

To simplify our discussion we assumed identical reference values of $n_e^0$ for
positive and negative streamers but microscopic models clearly show that
ionization is significantly stronger in positive streamers.  An example
can be found e.g. in ref.~\cite{Ihaddadene2015/GeoRL}, where the difference
is close to a factor 10.  From this we conclude that although our model
does not by itself explain the different velocities of streamers of opposite
polarities, it can account for this difference by an 
appropriate selection of the parameter $n_e^0$.

\subsection{Electrode potential}
The effect of a change in the potential applied to the electrode is easier to
explain.  A higher potential leads to higher electric fields and faster
propagation both in positive and negative streamers.  The effect is stronger
in positive streamers but this may be a consequence of our particular set
of parameters.

\subsection{Streamer radius}
Focusing now on the effect of streamer radius, we see that in general
a larger radius leads to a lower electric field but faster propagation.
This is consistent with the observations of ref.~\citep{Briels2008/JPhD/1}, which
were reproduced numerically in ref.~\citep{Luque2008/JPhD}.  The reason for
this behaviour is that a larger radius implies a slower decay of the electric
field ahead of the streamer, which greatly favours the multiplication of 
electrons and thus the further advance of the streamer.  This is accounted
for in the velocity expression (\ref{naidis}) derived by Naidis, which,
for a fixed peak electric field, predicts a higher velocity for a larger
streamer radius.

But note that for a positive streamer, the velocity is almost unchanged
between the reference simulation and the simulation with twice the streamer
radius.  In this case, the speed increase due to widening is compensated
by the decrease due to a lower peak electric field.  Here we also run into
another limitation of our model: as a streamer evolves, its radius and peak
electric field are interrelated.  This means that in general in a real
streamer a larger radius does not neccesarily imply a lower electric field.
The experimental relationship between radius and velocity, which is better 
defined than in our simulations, may be explained if 
the peak field does not decrease substantially for wider streamers.
This again underlines what we consider the main missing element in the model:
a self-consistent evolution of streamer radius.

\subsection{Velocity}
As mentioned above, we multiplied the velocity resulting from 
expression (\ref{naidis}) by a factor $\beta$ 
to investigate the role of streamer velocity.
Unsurprinsingly, a larger $\beta$ leads to faster propagation.

However, in figures~\ref{fig:batch_positive} and \ref{fig:batch_negative} we
also see that artificially slowed-down streamers have a higher electric field
at their tips.  This illustrates the competing dynamics that take place
in a streamer: the electrostatic relaxation of the streamer body strives to 
transport charge towards the tip but the ongoing propagation acts against
this accumulation of charge.  When we slow down the propagation we allow more
charge to reach the streamer head where it creates a higher enhanced field.
A quickly propagating streamer partly avoids this accumulation and therefore
has a lower peak electric field.

\section{Conclussions}
\label{sect:conclussions}
The work presented here is a further step towards the objective of realistic
and predictive simulations of complete discharge coronas.  Although so far
limited to single streamers, the model that we described 
includes a more accurate charge transport than the model 
of ref.~\citep{Luque2014/NJPh}.  Nevertheless, 
the discussion in the preceding section highlights some of the tradeoffs that we
considered when designing our scheme:
\begin{enumerate}
  \item The model includes parameters that have to be manually tuned in order
    to reproduce experimental observations.  As we discussed above, to explain
    the differences between positive and negative streamers, we have
    to assume different values of $n_e^0$.  Our model therefore has 
    less predictive power than a microscopical simulation.  We consider this
    an unavoidable price to pay for a macroscopic model. 
    Similar limitations occur in almost all branches of Physics.  For example,
    electromagnetic macroscopic models require material properties,
    such as electric permittivity, that must be obtained from microscopic
    calculations or directly from measurements.
  \item A more troublesome limitation is the lack of a self-consistent evolution
    of the streamer radius.  We have mentioned this issue at several places
    in this work and, as we discussed in section~\ref{sect:micro}, an evolving
    streamer radius is neccesary to account for the streamer dynamics observed
    in microscopic simulations.  We believe that
    this outstanding topic of streamer physics deserves to be the subject of
    future work.
\end{enumerate}

Even with those limitations, the model that we presented can be extended to
more realistic models of corona discharges that incorporate the strongly 
inhomogeneous field and electron density in the corona interior.  In principle
our scheme can be generalized to many interacting streamers with arbitrary 
shapes.  To achive that, however, further numerical optimizations are required.

The final objective of this type of approach to streamer modeling is to couple
it with the progression of a leader in order to understand how 
sections of the streamer corona are heated and join the leader channel.
We hope that this article contributes to bring this objective 
somewhat closer.

\textit{Acknowledgments}.-
This work was supported by the European Research Council (ERC) under the European Union H2020 programme/ERC grant agreement 681257 and by the Spanish Ministry of Science and Innovation, MINECO under projects FIS2014-61774-EXP and ESP2015-69909-C5-2-R.

\newcommand{\nat}{Nature}
\newcommand{\pre}{Phys. Rev. E}
\newcommand{\prb}{Phys. Rev. B}
\bibliography{library}

\providecommand{\newblock}{}
\begin{thebibliography}{10}
\expandafter\ifx\csname url\endcsname\relax
  \def\url#1{{\tt #1}}\fi
\expandafter\ifx\csname urlprefix\endcsname\relax\def\urlprefix{URL }\fi
\providecommand{\eprint}[2][]{\url{#2}}

\bibitem{Luque2014/NJPh}
{Luque} A and {Ebert} U 2014 {\em New Journal of Physics\/} {\bf 16} 013039
  (\textit{Preprint} \eprint{1307.2378})

\bibitem{Lozanskii1975/SvPhU}
{Lozanski{\u i}} {\'E}~D 1975 {\em Soviet Physics Uspekhi\/} {\bf 18} 893

\bibitem{Kao2010/PhysD}
Kao C~Y, Brau F, Ebert U, Schaefer L and Tanveer S {2010} {\em {Phys. D}\/}
  {\bf {239}} {1542--1559} ISSN {0167-2789}

\bibitem{Arrayas2010/PhRvE}
{Array{\'a}s} M, {Fontelos} M~A and {Jim{\'e}nez} C 2010 {\em Phys. Rev. E\/}
  {\bf 81} 035401 (\textit{Preprint} \eprint{0910.3617})

\bibitem{Ebert2011/Nonli/1}
{Ebert} U, {Brau} F, {Derks} G, {Hundsdorfer} W, {Kao} C~Y, {Li} C, {Luque} A,
  {Meulenbroek} B, {Nijdam} S, {Ratushnaya} V, {Sch{\"a}fer} L and {Tanveer} S
  2011 {\em Nonlinearity\/} {\bf 24} 1

\bibitem{Meulenbroek2004/PhRvE}
{Meulenbroek} B, {Rocco} A and {Ebert} U 2004 {\em Phys. Rev. E\/} {\bf 69}
  067402 (\textit{Preprint} \eprint{physics/0305112})

\bibitem{Meulenbroek2005/PhRvL}
{Meulenbroek} B, {Ebert} U and {Sch{\"a}fer} L 2005 {\em Phys. Rev. Lett.\/}
  {\bf 95} 195004 (\textit{Preprint} \eprint{nlin/0507019})

\bibitem{Arrayas2012/PhRvE}
{Array{\'a}s} M, {Fontelos} M~A and {Kindel{\'a}n} U 2012 {\em Phys. Rev. E\/}
  {\bf 86} 066407 (\textit{Preprint} \eprint{1203.6790})

\bibitem{Brau2008/PhRvE/1}
{Brau} F, {Luque} A, {Meulenbroek} B, {Ebert} U and {Sch{\"a}fer} L 2008 {\em
  Phys. Rev. E\/} {\bf 77} 026219 (\textit{Preprint} \eprint{0707.1402})

\bibitem{Brau2009/PhRvE}
{Brau} F, {Luque} A, {Davidovitch} B and {Ebert} U 2009 {\em Phys. Rev. E\/}
  {\bf 79} 066211 (\textit{Preprint} \eprint{0901.1916})

\bibitem{Brau2008PhRvE/2}
{Brau} F, {Davidovitch} B and {Ebert} U 2008 {\em \pre\/} {\bf 78} 056212

\bibitem{Arrayas2011/PhRvE}
{Array{\'a}s} M and {Fontelos} M~A 2011 {\em Phys. Rev. E\/} {\bf 84} 026404
  (\textit{Preprint} \eprint{1103.0404})

\bibitem{Niemeyer1984/PhRvL}
{Niemeyer} L, {Pietronero} L and {Wiesmann} H~J 1984 {\em Phys. Rev. Lett.\/}
  {\bf 52} 1033

\bibitem{Pasko2000/GeoRL}
{Pasko} V~P, {Inan} U~S and {Bell} T~F 2000 {\em Geophys. Res. Lett.\/} {\bf
  27} 497

\bibitem{Akyuz2003/JElec}
Akyuz M, Larsson A, Cooray V and Strandberg G 2003 {\em J. Electrost.\/} {\bf
  59} 115 -- 141 ISSN 0304-3886

\bibitem{Luque2012/JCoPh}
{Luque} A and {Ebert} U 2012 {\em J. Comput. Phys.\/} {\bf 231} 904

\bibitem{Dhali1985/PhRvA}
{Dhali} S~K and {Williams} P~F 1985 {\em Phys. Rev. A\/} {\bf 31} 1219

\bibitem{Pancheshnyi2003/JPhD}
{Pancheshnyi} S~V and {Starikovskii} A~Y 2003 {\em J. Phys. D\/} {\bf 36} 2683

\bibitem{Montijn2006/JCoPh}
{Montijn} C, {Hundsdorfer} W and {Ebert} U 2006 {\em J. Comput. Phys.\/} {\bf
  219} 801 (\textit{Preprint} \eprint{physics/0603070})

\bibitem{Luque2007/ApPhL}
{Luque} A, {Ebert} U, {Montijn} C and {Hundsdorfer} W 2007 {\em Appl. Phys.
  Lett.\/} {\bf 90} 081501 (\textit{Preprint} \eprint{physics/0609247})

\bibitem{Babaeva2009/PSST/1}
{Babaeva} N~Y and {Kushner} M~J 2009 {\em Plasma Sour. Sci. Technol.\/} {\bf
  18} 035009

\bibitem{Luque2010/GeoRL}
{Luque} A and {Ebert} U 2010 {\em Geophys. Res. Lett.\/} {\bf 37} L06806

\bibitem{Dujko2011/JaJAP}
{Dujko} S, {Ebert} U, {White} R~D and {Petrovi{\'c}} Z~L 2011 {\em Japanese
  Journal of Applied Physics\/} {\bf 50} 08JC01

\bibitem{Aleksandrov1996/JPhD}
{Aleksandrov} N~L and {Bazelyan} E~M 1996 {\em J. Phys. D\/} {\bf 29} 740

\bibitem{Jackson1975/book}
Jackson J 1975 {\em Classical electrodynamics\/} (New York, USA: John Wiley and
  sons) ISBN 9780471431329

\bibitem{Kossyi1992/PSST}
{Kossyi} I~A, {Kostinsky} A~Y, {Matveyev} A~A and {Silakov} V~P 1992 {\em
  Plasma Sour. Sci. Technol.\/} {\bf 1} 207

\bibitem{Aleksandrov1999/PSST}
{Aleksandrov} N~L and {Bazelyan} E~M 1999 {\em Plasma Sour. Sci. Technol.\/}
  {\bf 8} 285

\bibitem{Pancheshnyi2013/JPhD}
{Pancheshnyi} S 2013 {\em J. Phys. D\/} {\bf 46} 155201

\bibitem{Gallimberti1979/JPhys}
{Gallimberti} I 1979 {\em Journal de Physique\/} {\bf 40} 193

\bibitem{Dhali1987/JAP}
{Dhali} S~K and {Williams} P~F 1987 {\em J. Appl. Phys.\/} {\bf 62} 4696

\bibitem{Viehland1995/ADNDT}
{Viehland} L~A and {Mason} E~A 1995 {\em At. Data Nucl. Data Tables\/} {\bf 60}
  37

\bibitem{Pancheshnyi2012/CP}
{Pancheshnyi} S, {Biagi} S, {Bordage} M~C, {Hagelaar} G~J~M, {Morgan} W~L,
  {Phelps} A~V and {Pitchford} L~C 2012 {\em Chem. Phys.\/} {\bf 398} 148

\bibitem{Wissdorf2013/JASMS}
{Wissdorf} W, {Seifert} L, {Derpmann} V, {Klee} S, {Vautz} W and {Benter} T
  2013 {\em Journal of The American Society for Mass Spectrometry\/} {\bf 24}
  632

\bibitem{Naidis2009/PhRvE}
{Naidis} G~V 2009 {\em Phys. Rev. E\/} {\bf 79} 057401

\bibitem{Luque2008/PhRvL}
{Luque} A, {Ebert} U and {Hundsdorfer} W 2008 {\em Phys. Rev. Lett.\/} {\bf
  101} 075005 (\textit{Preprint} \eprint{0712.2774})

\bibitem{Luque2008/JPhD}
{Luque} A, {Ratushnaya} V and {Ebert} U 2008 {\em J. Phys. D\/} {\bf 41} 234005
  (\textit{Preprint} \eprint{0804.3539})

\bibitem{Luque2009/NatGe}
{Luque} A and {Ebert} U 2009 {\em Nature Geoscience\/} {\bf 2} 757

\bibitem{Luque2011/GeoRL}
{Luque} A and {Gordillo-V{\'a}zquez} F~J 2011 {\em Geophys. Res. Lett.\/} {\bf
  38} L04808

\bibitem{Luque2016/JGRA/temp}
{Luque} A, {Stenbaek-Nielsen} H~C, {Mc{H}arg} M~G and Haaland R~K 2016 {\em J.
  Geophys. Res. (Space Phys)\/} {\bf 121}

\bibitem{Hagelaar2005/PSST}
{Hagelaar} G~J~M and {Pitchford} L~C 2005 {\em Plasma Sour. Sci. Technol.\/}
  {\bf 14} 722

\bibitem{Phelps1985/PhRvA}
{Phelps} A~V and {Pitchford} L~C 1985 {\em Phys. Rev. A\/} {\bf 31} 2932

\bibitem{Ihaddadene2015/GeoRL}
{Ihaddadene} M~A and {Celestin} S 2015 {\em Geophys. Res. Lett.\/} {\bf 42}
  5644

\bibitem{Briels2008/JPhD/1}
{Briels} T~M~P, {Kos} J, {Winands} G~J~J, {van Veldhuizen} E~M and {Ebert} U
  2008 {\em J. Phys. D\/} {\bf 41} 234004 (\textit{Preprint}
  \eprint{0805.1376})

\end{thebibliography}

\end{document}